\documentclass[pra,twocolumn,floatfix,superscriptaddress]{revtex4}

\usepackage{graphicx}
\usepackage{color}
\usepackage[normalem]{ulem}
\usepackage{braket}
\usepackage{amsmath}
\usepackage{verbatim}
\usepackage{amssymb}

\newcommand{\be}{\begin{equation}}
\newcommand{\ee}{\end{equation}}
\newcommand{\ba}{\begin{eqnarray}}
\newcommand{\ea}{\end{eqnarray}}
\newcommand{\nn}{\nonumber \\}
\def\p{\partial}
\def\tr{\text{Tr}}

\begin{document}

\title{Quantum Boltzmann Machine}

\author{Mohammad H. Amin }
\affiliation{D-Wave Systems Inc., 3033 Beta Avenue, Burnaby BC
Canada V5G 4M9}
\affiliation{Department of Physics, Simon Fraser
University, Burnaby, BC Canada V5A 1S6}
\author{Evgeny Andriyash}
\affiliation{D-Wave Systems Inc., 3033 Beta Avenue, Burnaby BC
Canada V5G 4M9}
\author{Jason Rolfe}
\affiliation{D-Wave Systems Inc., 3033 Beta Avenue, Burnaby BC
Canada V5G 4M9}
\author{Bohdan Kulchytskyy}
\affiliation
{Department of Physics and Astronomy, University of Waterloo, 200 University Avenue West
 Waterloo, Ontario, Canada N2L 3G1}
\author{Roger Melko}
\affiliation
{Department of Physics and Astronomy, University of Waterloo, 200 University Avenue West
 Waterloo, Ontario, Canada N2L 3G1}
\affiliation{Perimeter Institute for Theoretical Physics, Waterloo, Ontario, N2L 2Y5, Canada}

\begin{abstract}

Inspired by the success of Boltzmann Machines based on classical Boltzmann distribution, we propose a new machine learning approach based on quantum Boltzmann distribution of a transverse-field Ising Hamiltonian. Due to the non-commutative nature of quantum mechanics, the training process of the Quantum Boltzmann Machine (QBM) can become nontrivial.  We circumvent the problem by introducing bounds on the quantum probabilities. This allows us to train the QBM efficiently by sampling. We show examples of QBM training with and without the bound, using exact diagonalization, and compare the results with classical Boltzmann training. We also discuss the possibility of using quantum annealing processors like D-Wave for QBM training and application.

\end{abstract}

\maketitle

\section{Introduction}

Machine learning is a rapidly growing field in computer science with applications in computer vision, voice recognition, medical diagnosis, spam filtering, search engines, etc.\cite{MLScience} Machine learning algorithms operate by constructing a model with parameters that can be determined (learned) from a large amount of example inputs, called {\em training set}. The trained model can then make predictions about unseen data. The ability to do so is called {\em generalization}. This could be, for example, detecting an object, like a cat, in an image or recognizing a command from a voice input. One approach to machine learning is probabilistic modeling in which the probability distribution of the data ($P^{\rm data}_{\bf v}$ for a given state ${\bf v}$) is approximated based upon a finite set of samples. If the process of training is successful, the learned distribution $P_{\bf v}$ has enough resemblance to the actual distribution of the data, $P^{\rm data}_{\bf v}$, such that it can make correct predictions about unseen situations. Depending upon the details of the distributions and the approximation technique, machine learning can be used to perform classification, clustering, collaborative filtering, compression, denoising, inpainting, or a variety of other algorithmic tasks \cite{Bishop}.

The possibility of using quantum computation for machine learning has been considered theoretically for both gate model \cite{Lloyd13,Lloyd14,Weibe14} and quantum annealing \cite{Neven08,Neven08b,Neven09,Pudenz11,deFreitas11,Denchev12,Dumoulin13,Babbush14} schemes. With the development of quantum annealing processors \cite{Johnson11}, it has become possible to test machine learning ideas with an actual quantum hardware \cite{Adachi15,Alejandro15}. In all of the above works, however, the quantum processor is considered as a means to provide fast solutions to an otherwise classical problem. In other words, the model stays classical and quantum mechanics is only used to facilitate the training. In this work, we propose a quantum probabilistic model for machine learning based on Boltzmann distribution of a quantum Hamiltonian, therefore, a Quantum Boltzmann Machine (QBM). As we shall see, in our approach, the quantum nature of the processor is exploited {\em both in the model and in the training process}.

The Boltzmann machine (BM) is a classic machine learning technique, and serves as the basis of powerful deep learning models such as deep belief networks and deep Boltzmann machines \cite{Hinton1983, Hinton2006, Salakhutdinov2009}.
It comprises a probabilistic network of binary units with a quadratic energy function. In principle, one could consider more general energy functions to bring in more flexibility \cite{Sejnowski1986, Ranzato2010, Memisevic2010}, but training can become impractical and generalization suffers as the number of parameters grows. A BM commonly consists of visible and hidden binary units, which we jointly denote by $z_a$, $a=1,...,N$, where $N$ is the total number of units.  To maintain consistency with the standard notation in quantum mechanics, we use $z_a \in \{-1, +1\}$, rather than $z_a \in \left\{0, 1 \right\}$; the corresponding probability distributions are identical up to a linear transformation of their parameters.  To distinguish the visible and hidden variables, we use the notation $z_a  =(z_\nu, z_i)$, with index $\nu$ for visible variables and $i$ for hiddens. We also use vector notations ${\bf v}$, ${\bf h}$, and ${\bf z} = ({\bf v},{\bf h})$ to represent states of visible, hidden, and combined units, respectively. In physics language, the quadratic energy function over binary units $z_a$ is referred to as Ising model with the energy function:
 \ba
 E_{\bf z} = - \sum_{a} b_a z_a - \sum_{a,b} w_{ab} z_a z_b. \label{Ising}
 \ea
The dimensionless parameters $b_a$ and $w_{ab}$ are tuned during the training \footnote{In physical systems, Hamiltonian parameters have unit of energy. We normalize these parameters by  $k_B T \equiv \beta^{-1}$, where $T$ is temperature and $k_B$ is the Boltzmann constant; we absorb $\beta$ into the parameters.}. In equilibrium, the probability of observing a state ${\bf v}$ of the visible variables is given by the Boltzmann distribution summed over the hidden variables:
 \be\label{PV}
 P_{\bf v} = Z^{-1} \sum_{\bf h} e^{- E_{\bf z}},\ \qquad Z = \sum_{\bf z} e^{-E_{\bf z}},
 \ee
called {\em marginal} distribution. Our goal is to determine Hamiltonian parameters, $\theta {\in} \{b_a,w_{ab} \}$, such that $P_{\bf v}$ becomes as close as possible to $P_{\bf v}^{\rm data}$ defined by the training set. To achieve this, we need to maximize the average log-likelihood, or equivalently minimize the average negative log-likelihood defined by
 \be \label{NLL}
 {\cal L} = -\sum_{\bf v} P_{\bf v}^{\rm data} \log P_{\bf v},
 \ee
which for the probability distribution (\ref{PV}) is
 \be
 {\cal L} = -\sum_{\bf v} P_{\bf v}^{\rm data} \log {\sum_{\bf h} e^{-E_{\bf z}} \over \sum_{\bf z'} e^{-E_{\bf z'}}}. \label{Lclassical}
 \ee
The minimization can be done using gradient decent technique. In each iteration, the parameter $\theta$ is changed by a small step in the direction opposite to the gradient:
 \be
 \delta \theta = -\eta \p_\theta {\cal L},
 \ee
where the learning rate, $\eta$, controls the step sizes. An important requirement for applicability of the gradient decent technique is the ability to calculate the gradients $\p_\theta {\cal L}$ efficiently.
Using (\ref{Lclassical}), we have
 \ba
 \p_\theta {\cal L}
 &=&   \sum_{\bf v} P_{\bf v}^{\rm data} {\sum_{\bf h} \p_\theta E_{\bf z} e^{-E_{\bf z}} \over {\sum_{\bf h} e^{-E_{\bf z}} }} - {\sum_{\bf z} \p_\theta E_{\bf z} e^{-E_{\bf z}} \over \sum_{\bf z} e^{-E_{\bf z}}} \nn
 &=&  \overline{\langle \p_\theta E_{\bf z}\rangle_{\bf v}}
 - \langle \p_\theta E_{\bf z} \rangle,
 \ea
where $\langle ... \rangle$  and $\langle ... \rangle_{\bf v}$ are Boltzmann averages with free and fixed visible variables, respectively, and $\overline{\langle ... \rangle_{\bf v}} \equiv \sum_{\bf v} P_{\bf v}^{\rm data} \langle ... \rangle_{\bf v}$ denotes double averaging. Fixing visible variables to the data is usually called {\em clamping}. Using (\ref{Ising}) for $E_{\bf z}$, we obtain
 \ba
 \delta b_a &=&  \eta \left( \overline{\langle z_a \rangle_{\bf v}} - \langle z_a \rangle \right),\\
 \delta w_{ab} &=&  \eta \left( \overline{\langle z_az_b \rangle_{\bf v}} - \langle z_az_b \rangle \right).
 \ea
The gradient steps are expressed in terms of differences between the clamped (i.e., fixed ${\bf v}$) and unclamped averages. These two terms are sometimes called {\em positive} and {\em negative} phases. Since the averages can be estimated using sampling, the process of gradient estimation can be done efficiently provided that we have an efficient way of performing sampling.

\begin{figure}[t]
   \includegraphics[trim = 20mm 0mm 10mm 30mm, width=7cm]{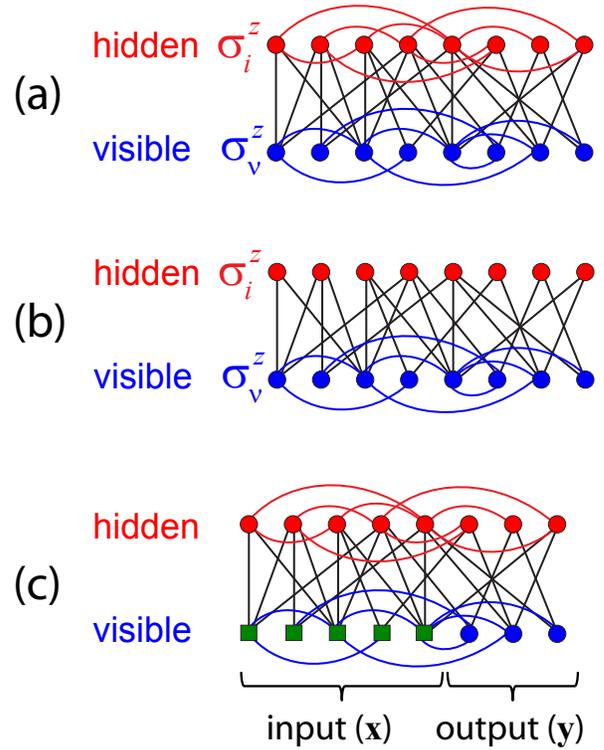}
\caption{(a) An example of a quantum Boltzmann machine with visible (blue) and hidden (red) qubits. (b) A restricted quantum Boltzmann machine with no lateral connection between the hidden variables. (c) Discriminative learning with QBM. The (green) squares represent classical input ${\bf x}$, which are not necessarily binary numbers. The input applies energy biases to the hidden and output qubits according to the coupling coefficients represented by solid lines.}
    \label{fig1}
\end{figure}

\section{Quantum Boltzmann machine}\label{sec:QBM}

We now replace the classical spins or bits in (\ref{Ising}) with quantum bits (qubits).  The mathematics of quantum mechanics is based on matrices (operators) with dimensionality equal to the number of possible states $(2^N)$. This in contrast to vectors with dimensionality equal to the number of variables $(N)$ used in common machine learning techniques. For instance, instead of the energy function (\ref{Ising}), one considers a $2^N{\times}2^N$ diagonal matrix, called the Hamiltonian:
\ba\label{Hdiag}
 H = - \sum_{a} b_a \sigma^z_a - \sum_{a,b} w_{ab} \sigma^z_a \sigma^z_b.
 \ea
This Hamiltonian is constructed in such a way that its diagonal elements are energy values (\ref{Ising}) corresponding to all $2^N$ binary states ${\bf z}$ ordered  lexicographically. To generate such a Hamiltonian,  we replace $z_a$ in (\ref{Ising}) with $2^N{\times}2^N$ matrix
 \be
 \sigma^z_a \equiv \overbrace{I\otimes ... \otimes I}^{a-1}\otimes \sigma_z \otimes \overbrace{I \otimes... \otimes I}^{N-a} \label{sigma_def}
 \ee
where $\otimes$ means tensor product (sometimes called Kronecker or outer product) and
 \be
 I = \left( \begin{array}{cc} 1 & 0 \\ 0 & 1 \end{array} \right), \qquad
 \sigma_z = \left( \begin{array}{cc} 1 & 0 \\ 0 & -1 \end{array} \right).
 \ee
Every element in (\ref{sigma_def}) is an identity matrix ($I$) except the $a$-th element which is a Pauli matrix ($\sigma_z$). Equation (\ref{Ising}) will therefore be replaced by the diagonal Hamiltonian where $b_a$ and $w_{ab}$ are still scalars.
Fig.~\ref{fig1}a shows an example of such a model with visible and hidden qubits depicted as blue and red circles, respectively. We represent eigenstates of this Hamiltonian by $\ket{\bf v,h}$, where again ${\bf v}$ and ${\bf h}$ denote visible and hidden variables, respectively.

We can now define matrix exponentiation through Taylor expansion, $e^{-H} = \sum_{k=0}^\infty \frac{1}{k!} (-H)^k$. For a diagonal Hamiltonian, $e^{-H}$ is a diagonal matrix with its $2^N$ diagonal elements being $e^{-E_{\bf z}}$ corresponding to all the $2^N$ states. With the partition function given by $Z=\tr[e^{- H}]$ (c.f. (\ref{PV})), we define the density matrix as
\be\label{eq:rho}
\rho = Z^{-1}e^{- H}.
\ee
The diagonal elements of $\rho$ are therefore Boltzmann probabilities of all the $2^N$ states.
For a given state $\ket{\bf v}$ of the visible variables, we can obtain the marginal Boltzmann probability $P_{\bf v}$ by tracing over the hidden variables
 \be
 P_{\bf v} = \tr[ \Lambda_{\bf v}\rho], \label{PvQ}
 \ee
where $\Lambda_{\bf v}$ limits the trace only to diagonal terms that correspond to the visible variables being in state ${\bf v}$. Thus, $\Lambda_{\bf v}$ is a diagonal matrix with diagonal elements being either 1, when the visibles are in state  ${\bf v}$, or 0 otherwise. In operator notation, we write
 \be
 \Lambda_{\bf v} = \ket{\bf v}\bra{\bf v} \otimes {\cal I}_{\bf h}, \label{Ldav}
 \ee
where ${\cal I}_{\bf h}$ is the identity matrix acting on the hidden variables, and
 \be
 \ket{\bf v}\bra{\bf v} \equiv \prod_\nu \left({1+{\bf v}_\nu \sigma^z_\nu \over 2}\right)
 \ee
is a projection operator in the subspace of visible variables. Equations (\ref{PV}) and (\ref{PvQ}) are equivalent when the Hamiltonian and therefore the density matrix are diagonal, but (\ref{PvQ}) also holds for non-diagonal matrices.

We can now add a transverse field to the Ising Hamiltonian by introducing non-diagonal matrices
 \ba
 \sigma^x_a \equiv \overbrace{I\otimes ... \otimes I}^{a-1}\otimes \sigma_x \otimes \overbrace{I \otimes... \otimes I}^{N-a}, \qquad
 \sigma_x = \left( \begin{array}{cc} 0 & 1 \\ 1 & 0 \end{array} \right), \nonumber
 \ea
which represent transverse components of spin. The transverse Ising Hamiltonian is then written as
 \ba\label{eq:Hamiltonian}
 H = - \sum_{a} \Gamma_a \sigma^x_a - \sum_{a} b_a \sigma^z_a - \sum_{a,b} w_{ab} \sigma^z_a \sigma^z_b \label{HTI}
 \ea
Every eigenstate of $H$ is now a superposition in the computation basis made of the classical states $\ket{\bf v,h}$. As the probabilistic model for QBM, we use quantum Boltzmann distribution with the density matrix (\ref{eq:rho}), which now has off-diagonal elements. In each measurement the states of the qubits are read out in the $\sigma_z$-basis and the outcome will be a classical value $\pm1$. Because of the statistical nature of quantum mechanics, after each measurement a classical output ${\bf v}$ will appear for the visible variables with the probability $P_{\bf v}$ given by (\ref{PvQ}).

To train a QBM, we change the parameters $\theta$ such that the probability distributions $P_{\bf v}$ becomes close to $P_{\bf v}^{\rm data}$ of the input data. This is achieved by minimizing the negative log-likelihood, which from (\ref{NLL}), (\ref{eq:rho}), and (\ref{PvQ}) is
 \be
 {\cal L} = -\sum_{\bf v} P_{\bf v}^{\rm data} \log {\tr[ \Lambda_{\bf v}e^{- H}] \over \tr[e^{- H}]}. \label{LLambda}
 \ee
The gradient of ${\cal L}$ is given by
 \be
 \p_\theta {\cal L}
 =  \sum_{\bf v} P_{\bf v}^{\rm data} \left({\tr[\Lambda_{\bf v}\p_\theta e^{- H}] \over \tr[\Lambda_{\bf v}e^{- H}]} - {\tr[\p_\theta e^{- H}] \over \tr[e^{- H}]}\right). \label{pL}
 \ee
Once again, we hope to be able to estimate the gradients efficiently using sampling. However, since $H$ and $\partial_\theta H$ are now matrices that do not commute, we have $\partial_\theta e^{-H} \ne - e^{-H} \partial_\theta H$ and therefore we don't trivially obtain expectations of $\partial_\theta H$ as in the classical case. Writing $e^{- H} = [e^{-\delta \tau H}]^n$, where $\delta \tau \equiv 1/n$, we have
\ba
\p_\theta e^{-H} = \sum_{m=1}^n  e^{- m\delta \tau H} \left( - \p_\theta H \delta \tau \right)  e^{-(n-m) \delta \tau H}.
\ea
Introducing imaginary time $\tau \equiv m\delta\tau$, in the limit of $n\to \infty$, we obtain
\ba\label{eq:variation}
&& \p_\theta e^{-H} = - \int_0^1 d\tau e^{-\tau H}   \p_\theta H   e^{(\tau-1) H}.
\ea
Tracing over both sides and using permutation property of the trace, we find
\be
\tr[\p_\theta e^{- H}] = - \tr[\p_\theta H e^{- H}],
\ee
which is the same as the classical relation. Plugging this into the second term of  (\ref{pL}) gives
\be
{\tr[\p_\theta e^{- H}] \over \tr[e^{- H}]} = -\langle \p_\theta H\rangle,
\ee
where $\langle ... \rangle \equiv \tr [\rho ...]$ denotes Boltzmann averaging. This term can be estimated by sampling from the distribution (\ref{eq:rho}). However, the first term in  (\ref{pL}),
\be
{\tr[\Lambda_{\bf v}\p_\theta e^{- H}] \over \tr[\Lambda_{\bf v}e^{- H}]} = - \int_0^1 d t  {\tr[\Lambda_{\bf v} e^{-t H}  \p_\theta H e^{-(1-t) H}] \over \tr[\Lambda_{\bf v}e^{- H}]},
\ee
cannot be estimated using sampling. This renders the training of a QBM inefficient and basically impractical for large system. A work around for this problem is to introduce a properly defined upper-bound for ${\cal L}$ and minimize it, as we shall discuss below. We call this approach bound-based QBM (bQBM). Minimizing a bound on the negative log-likelihood is a common approach in machine learning.

\subsection{Bound-based QBM}

One can define a lower bound for the probabilities using Golden-Thompson inequality \cite{Golden65,Thompson65}:
 \be
 \tr[e^A e^B] \ge \tr[e^{A+B}], \label{GT}
 \ee
which holds for any hermitian matrices $A$ and $B$. We can therefore write
 \be
 P_{\bf v} = {\tr[e^{- H} e^{\ln \Lambda_{\bf v}}] \over \tr[e^{- H}]}
 \geq {\tr[e^{- H + \ln \Lambda_{\bf v}}] \over \tr[e^{- H}]}.
 \ee
Introducing a new Hamiltonian:
 \be
 H_{\bf v} = H - \ln \Lambda_{\bf v},
 \ee
we can write
 \be
 P_{\bf v} \geq {\tr[e^{- H_{\bf v}}] \over \tr[e^{- H}]} . \label{Pmeasineq}
 \ee
Notice that $H_{\bf v}$ has an infinite energy penalty for any state of the visible qubits that is different from $\ket{\bf v}$. Therefore, for any practical purposes,
 \be\label{eq:H_clamped}
 H_{\bf v} \equiv H(\sigma^x_\nu =0, \sigma^z_\nu = {\bf v}_\nu).
 \ee
This is a {\em clamped} Hamiltonian because every visible qubit $\sigma^z_\nu$ is clamped to its corresponding classical data value ${\bf v}_\nu$.

From (\ref{LLambda}) and (\ref{Pmeasineq}) it follows that
 \be\label{eq:bound}
 {\cal L} \leq \tilde{\cal L} \equiv -\sum_{\bf v} P_{\bf v}^{\rm data} \log {\tr[e^{- H_{\bf v}}] \over \tr[e^{- H}]}.
 \ee
Instead of minimizing ${\cal L}$, we now minimize its upper bound $\tilde{\cal L}$ using the gradient
 \ba\label{eq:bound_gradient}
 \p_\theta \tilde{\cal L}
 &=&   \sum_{\bf v} P_{\bf v}^{\rm data} \left({\tr[e^{- H_{\bf v}}\p_\theta H_{\bf v}] \over \tr[e^{- H_{\bf v}}]} - {\tr[e^{- H}\p_\theta H] \over \tr[e^{- H}]}\right), \nn
 &=&  \left( \overline{\langle \p_\theta H_{\bf v} \rangle_{\bf v}} - \langle \p_\theta H \rangle \right),
 \ea
where
 \ba
 \overline{\langle ... \rangle_{\bf v}} = \sum_{\bf v} P_{\bf v}^{\rm data} \langle ... \rangle_{\bf v}
  = \sum_{\bf v} P_{\bf v}^{\rm data} {\tr e^{- H_{\bf v}} ... \over  \tr e^{- H_{\bf v}}}.
 \ea
Taking $\theta$ to be $b_a$, $w_{ab}$, and using $\delta \theta = -\eta \p_\theta \tilde{\cal L}$, we obtain
 \ba\label{eq:qb_gradient}
 \delta b_a &=&  \eta \left( \overline{\langle \sigma^z_a \rangle_{\bf v}} - \langle \sigma^z_a \rangle \right),\\
\delta w_{ab} &=&  \eta \left( \overline{\langle \sigma^z_a\sigma^z_b \rangle_{\bf v}} - \langle \sigma^z_a\sigma^z_b \rangle \right). \label{eq:qw_gradient}
 \ea
Again the gradient steps are expressed in terms of differences between the unclamped and clamped averages, $\langle ... \rangle$ and $\langle ... \rangle_{\bf v}$, which can be obtained by sampling from a Boltzmann distribution with Hamiltonians $H$ and $H_{\bf v}$, respectively. In Section \ref{sec:Examples}, we give examples of training QBM and compare the results of minimizing ${\cal L}$ using (\ref{pL}) with minimizing its upper bound $\tilde {\cal L}$  using (\ref{eq:bound_gradient}).

One may also attempt to train $\Gamma_a$ using the upper bound $\tilde {\cal L}$. From (\ref{eq:bound_gradient}) we obtain
 \be\label{eq:Delta_gradient}
 \delta \Gamma_a = \eta \left( \overline{\langle \sigma^x_a \rangle_{\bf v}} - \langle \sigma^x_a \rangle \right).
 \ee
There are a few problems with using (\ref{eq:Delta_gradient}) to train $\Gamma_a$. First of all, one cannot calculate $\langle \sigma^x_a \rangle$ by sampling in $\sigma^z_a$ basis. Therefore, measurement in the $\sigma^x_a$ basis is needed to estimate $\langle \sigma^x_a \rangle$. Moreover, the first term in (\ref{eq:Delta_gradient}) is always zero for visible variables, i.e., $\overline{\langle \sigma^x_\nu \rangle_{\bf v}}=0, \ \forall \nu$. Since $\langle \sigma^x_\nu \rangle >0$ for positive $\Gamma_\nu$, $\delta \Gamma_\nu$ will always be negative, which means $\Gamma_\nu\to 0$ for all visible variables. This is inconsistent with what we obtain when we train $\Gamma_\nu$ using the exact gradient (\ref{pL}). Therefore, vanishing $\Gamma_\nu$ is an artifact of the upper bound minimization. In other words, we cannot learn the transverse field using the upper bound. One may still train the transverse field using the exact log-likelihood, but it becomes quickly inefficient as the size of the QBM grows.

\subsection{Restricted QBM}\label{subsec:RQBM}

So far we haven't imposed any restrictions on the connectivity between visible and hidden qubits or lateral connectivity among visible or hidden qubits. We note that calculation of the first term in (\ref{eq:qb_gradient}) and (\ref{eq:qw_gradient}), sometimes called positive phase, requires sampling from distributions with clamped Hamiltonians (\ref{eq:H_clamped}). This sampling can become computationally expensive for a large data set, because it has to be done for every input data element. If we {\em restrict} our QBM to have no lateral connectivity in the hidden layer (see Fig.~\ref{fig1}b), the hidden qubits become uncoupled in the positive phase and the calculations can be carried out exactly. We can write the clamped Hamiltonian (\ref{eq:H_clamped}) as
\be
H_{\bf v}  = - \sum_{i} \left( \Gamma_i \sigma^x_i  + b_i^{\rm eff}({\bf v}) \sigma^z_i  \right),
\ee
where $b_i^{\rm eff}({\bf v})  = b_i +  \sum_{\nu} w_{i\nu} {\bf v}_\nu$.  Expectations $\langle \sigma^z_i\rangle_{\bf v}$ entering (\ref{eq:qb_gradient}) can be computed exactly:
 \ba\label{eq:hidden_exp}
 \langle \sigma^z_i\rangle_{\bf v} =  \frac{b_i^{\rm eff}}{D_i} \tanh D_i,
\ea
where $D_i = \sqrt{\Gamma_i^2 +(b_i^{\rm eff})^2 }$. Notice that (\ref{eq:hidden_exp}) reduces to the classical RBM expression,
 \ba\label{eq:classical_exp}
 \langle \sigma^z_i\rangle_{\bf v} = \tanh b_i^{\rm eff},
 \ea
in the limit $\Gamma_i \to 0$. We emphasize that unlike the classical RBM, in which there are no lateral connections in both hidden and visible layers (for contrastive divergence techniques to work), we only require their absence in the hidden layer, usually called semi-restricted Boltzmann machine \cite{semirestricted}. In Section \ref{sec:Examples} we give an example of training RQBM and illustrate the importance of using (\ref{eq:hidden_exp}) instead of their classical limit.

\section{Supervised learning}

One important application of machine learning is classification in which a category (label) is assigned to each data point. For example, in spam detection the goal is to determine which of the two labels, ``spam'' or ``not spam'', should be assigned to a given text. The process of inferring a functional relation between input and label from a set of labeled data is called {\em supervised} learning. Denoting the feature vector (input) by ${\bf x}$  and label (output) by ${\bf y}$, the problem is to infer a function $g({\bf x}): {\bf x} \to {\bf y}$ from the set of labeled data $({\bf x}_i, {\bf y}_i)$. In probabilistic approaches to this problem, which are of our main interest here, the output ${\bf y}$ that is most probable, subject to the input ${\bf x}$, is chosen as the label. Therefore, the function $g({\bf x})$ is determined by the conditional probability $P_{\bf y|x}$ of output given input
\be
g({\bf x}) = \arg \max_{\bf y} P_{\bf y|x}.
\ee
The end goal of training is to make $P_{\bf y|x}$ as close as possible to the conditional distribution of the data, $P^{\rm data}_{\bf y|x}$. Assuming that the data comes with a joint probability distribution $P_{\bf x,y}^{\rm data}$, we can write: $P_{\bf y|x}^{\rm data} = P_{\bf x,y}^{\rm data}/P_{\bf x}^{\rm data}$, where $P_{\bf x}^{\rm data} = \sum_{\bf y} P_{\bf x,y}^{\rm data}$ is the marginal distribution.

Two possible approaches to supervised learning are {\em discriminative} and {\em generative} learning \footnote{There are other techniques used for supervised learning, for example, when only a small fraction of the available data is labeled.}. In the discriminative approach, for each ${\bf x}$ we try to learn the conditional distribution $P^{\rm data}_{\bf y|x}$. If an input ${\bf x}$ appears in the training set with probability $P_{\bf x}^{\rm data}$, the loss function can be written as
\ba\label{eq:L_discr}
 {\cal L}_{\rm discr} &=& -\sum_{\bf x} P_{\bf x}^{\rm data} \sum_{\bf y} P_{\bf y|x}^{\rm data}\log P_{\bf y|x}, \nn
 &=& -\sum_{\bf x,y} P_{\bf x,y}^{\rm data} \log P_{\bf y|x}.
\ea
In the generative approach, on the other hand, we learn the joint probability distribution without separating input from output. The loss function is therefore:
\ba\label{eq:L_gen}
 {\cal L}_{\rm gen} &=& -\sum_{\bf x,y} P_{\bf x,y}^{\rm data} \log P_{\bf x,y}\nn
 &=& {\cal L}_{\rm discr} -\sum_{\bf x} P_{\bf x}^{\rm data} \log P_{\bf x}
\ea
where we have used $P_{\bf x,y} = P_{\bf y|x} P_{\bf x}$. Notice that the first term is just $ {\cal L}_{\rm discr}$ while the second term measures the difference between the probability distribution of the training set inputs and the marginal distribution $P_{\bf x}$. This second term is called cross-entropy and it is equal to KL-divergence, see Eq.~(\ref{eq:KL_def}), up to a constant. Now, we explore the possibility of applying QBM to both cases.

\subsection{Generative learning} \label{GenLearning}

Generative learning with loss (\ref{eq:L_gen}) can be done with the methods of Section \ref{sec:QBM} by treating input and output (${\bf x,y}$) jointly as the visible data $\bf v = [\bf{x, y}]$ in a QBM. At the end of training, the QBM provides samples with a joint probability $P_{\bf x,y}$ that is close to $P_{\bf x,y}^{\rm data}$. Therefore, the conditional probability
\ba\label{eq:conditional_prob}
P_{\bf y|x}  &=& {P_{\bf x,y} \over P_{\bf x}} = {\tr[\Lambda_{\bf x}  \Lambda_{\bf y} e^{- H}] \over \tr[ \Lambda_{\bf x} e^{- H}]}.
\ea
should also match $P_{\bf y|x}^{\rm data}$ as desired for supervised training. However, there is a problem when it comes to sampling from this conditional for a given $\bf x$. If the input $\bf x$ appears with a very small probability ($P_{\bf x}\ll 1$), it would require a large amount of samples from $P_{\bf x,y}$ and $P_{\bf x}$ to reliably calculate $P_{\bf y|x}$ using (\ref{eq:conditional_prob}).

In a classical BM, one can sample from the conditional distribution by clamping the input variables ${\bf x}$ to the data and sampling the output ${\bf y}$. To understand how that strategy would work for QBM, let us introduce a clamped Hamiltonian
\be
H_{\bf x}=H-\ln \Lambda_{\bf x}, \qquad
\Lambda_{\bf x} =  \ket{\bf x}\bra{\bf x} \otimes {\cal I}_{\bf y} \otimes {\cal I}_{\bf h},
\ee
which clamps the input qubits to ${\bf x}$. Here, ${\cal I}_{\bf y}$ and ${\cal I}_{\bf h}$ are identity matrices acting on the output and hidden variables respectively.
For classical Hamiltonians ($[H,\Lambda_{\bf x}]{=} 0$), we have
\be\label{eq:conditional_prob_clamped}
P_{\bf y|x}  = {\tr[\Lambda_{\bf y} e^{- H}e^{\ln \Lambda_{\bf x}}] \over \tr[e^{- H}e^{\ln \Lambda_{\bf x}}]}
= P_{\bf y|x}^{\rm clamped},
\ee
where
\be
P_{\bf y|x}^{\rm clamped} \equiv  {\tr[\Lambda_{\bf y} e^{- H_{\bf x}}] \over \tr[e^{- H_{\bf x}}]}.
\ee
This means for any $\bf x$, we can sample $P_{\bf y|x}^{\rm clamped}$ from $H_{\bf x}$ and that will give us $P_{\bf y|x}$ in an efficient way regardless of how small $P_{\bf x}$ is. For quantum Hamiltonians, when $[H,\Lambda_{\bf x}]{\neq} 0$, we know that $e^{- H}e^{\ln \Lambda_{\bf x}} \neq e^{- H_{\bf x}}$. Therefore, $P_{\bf y|x}^{\rm clamped}$ is not necessarily equal to $P_{\bf y|x}$ and there is no easy way to draw samples from $P_{\bf y|x}$.

One might still hope that the clamped distribution is not too far off from (\ref{eq:conditional_prob}) and can be used as an approximation $P^{\rm clamped}_{\bf y|x}  \approx P_{\bf y|x}$. As we shall see in an example in Sec.~\ref{sec:Examples}-C, this is not true in general.

\subsection{Discriminative learning}

In discriminative learning one distinguishes input from output during the training \cite{Bishop} and learns the conditional probability distribution using (\ref{eq:L_discr}). This can be done by clamping the input ${\bf x}$ in both positive and negative phases. Since the input is always clamped, its role is just to apply biases to the other variables and therefore we don't need to assign any qubits to the input (see Fig.~\ref{fig1}c). The Hamiltonian of the system for a particular state of the input, ${\bf x}$, is given by
 \be
 H_{\bf x} = -\sum_{a} [\Gamma_a \sigma^x_a + b^{\rm eff}_a({\bf x}) \sigma^z_a] - \sum_{a,b}
 w_{ab}\sigma^z_a\sigma^z_b, \label{Hx}
 \ee
where indices $a$ and $b$ range over both hidden and visible (output only) variables. Here, the input ${\bf x}$ provides a bias
 \be
 b^{\rm eff}_a({\bf x}) = b_a {+} \sum_\mu w_{a\mu}x_\mu
 \ee
to the $a$-th qubit, where $b_a$ and $w_{a\mu}$ are tunable parameters. Notice that $x_\mu$ does not need to be restricted to binary numbers, which can bring more flexibility to the supervised learning.

The probability of measuring an output state ${\bf y}$ once the input is set to state ${\bf x}$ is given by
 \ba
 P_{\bf y|x} = {\tr[\Lambda_{\bf y} e^{- H_{\bf x}}] \over \tr[e^{- H_{\bf x}}]}, \qquad
 \Lambda_{\bf y} =  {\cal I}_{\bf x} \otimes \ket{\bf y}\bra{\bf y} \otimes {\cal I}_{\bf h},
 \ea
where $H_{\bf x}$ is given by (\ref{Hx}) and ${\cal I}_{\bf x}$ is an identity matrix acting on the input variables. The negative log-likelihood is given by (\ref{eq:L_discr}). Using the same tricks as discussed in the previous section we can define a clamped Hamiltonian,
 \be
 H_{\bf x,y} = H_{\bf x} - \ln \Lambda_{\bf y},
 \ee
and show that
 \be
 P_{\bf y|x} \gtrsim {\tr [e^{- H_{\bf x,y}}] \over \tr [e^{- H_{\bf x}}]}.
 \ee
Again we introduce an upper bound $\tilde{\cal L}$ for the ${\cal L}$
 \be
 {\cal L}_{\rm discr} \leq  \tilde{\cal L}_{\rm discr} = -\sum_{\bf x,y} P_{\bf x,y}^{\rm data}  \log {\tr [e^{- H_{\bf x,y}}] \over \tr [e^{- H_{\bf x}}]}.
 \ee
The derivative of $\tilde{\cal L} $ with respect to a Hamiltonian parameter $\theta$ is given by
 \ba
 \p_\theta \tilde {\cal L}
 =  \overline{\langle \p_\theta H_{\bf x,y} \rangle_{\bf x,y}} - \overline{\langle \p_\theta H_{\bf x} \rangle_{\bf x}},
 \ea
where
 \ba
 \overline{\langle A \rangle_{\bf x}} &=& \sum_{\bf x} P_{\bf x}^{\rm data}{\tr [e^{- H_{\bf x}} A] \over  \tr [e^{- H_{\bf x}}]}, \\
 \overline{\langle A \rangle_{\bf x,y}} &=&\sum_{\bf x,y} P_{\bf x,y}^{\rm data}
 {\tr [e^{- H_{\bf x,y}} A] \over  \tr [e^{- H_{\bf x,y}}]}.
 \ea
The gradient descent steps in the parameter space are given by
 \ba
 \delta b_a &=&  \eta \left( \overline{\langle \sigma^z_a \rangle_{\bf x,y}} - \overline{\langle \sigma^z_a \rangle_{\bf x}} \right), \label{dhs} \\
 \delta w_{ab} &=&  \eta \left( \overline{\langle \sigma^z_a\sigma^z_b \rangle_{\bf x,y}} - \overline{\langle \sigma^z_a\sigma^z_b \rangle_{\bf x}} \right), \label{dJs} \\
 \delta w_{a\mu} &=&  \eta \left( \overline{\langle \sigma^z_a x_\mu \rangle_{\bf x,y}} - \overline{\langle \sigma^z_a x_\mu \rangle_{\bf x}} \right). \label{dJas}
 \ea
Notice that ${\bf x}$ is not only clamped in the positive phase (the first expectations), but also is clamped in the negative phase (the second expectations). The positive phase can still be done efficiently if we use RQBM (Fig.~\ref{fig1}c with no lateral connection among the hidden variables). The negative phase needs a more elegant sampling method. This can make the calculation of the gradient steps computationally expensive for large data sets, unless a very fast sampling method is available.

\section{Examples}\label{sec:Examples}

In this Section we describe a few toy examples illustrating the ideas described in the previous sections. In all examples studied, the training data was generated as a mixture of $M$ factorized distributions (modes), each peaked around a random point. Every mode ($k$) is constructed by randomly selecting a center point ${\bf s}^k = [s^k_1,s^k_2,...,s^k_N]$ with $s_i^k \in \{ \pm 1\}$ and using Bernoulli distribution: $p^{N-d^k_{\bf v}}(1-p)^{d^k_{\bf v}}$, where $p$ is the probability of qubit $\nu$ being aligned with $s_\nu^k$, and $d^k_{\bf v}$ is the Hamming distance between ${\bf v}$ and ${\bf s}^k$. The average probability distribution over $M$ such modes gives our data distribution
\be\label{eq:mixture_model}
P^{\rm data}_{\bf v} = {1 \over M} \sum_{k=1}^M p^{N-d^k_{\bf v}}(1-p)^{d^k_{\bf v}},
\ee
In all our examples, we choose $p=0.9$ and $M=8$.

To have a measure of the quality of learning, we subtract from ${\cal L}$ its minimum ${\cal L}_{\rm min} = -\sum_{\bf v} P_{\bf v}^{\rm data} \log P_{\bf v}^{\rm data}$, which happens when $P_{\bf v} = P_{\bf v}^{\rm data}$. The difference, commonly called Kullback-Leibler (KL) divergence:
 \be\label{eq:KL_def}
 \text{KL} = {\cal L} - {\cal L}_{\rm min} = \sum_{\bf v} P_{\bf v}^{\rm data}
 \log {P_{\bf v}^{\rm data} \over P_{\bf v}},
 \ee
is a non-negative number measuring the difference between the two distributions; KL = 0 if and only if the two distributions are identical.

\subsection{Fully visible model}

We start with a fully visible example to compare BM with QBM and evaluate the quality of the bound (\ref{eq:bound}) by training bQBM. We consider a fully connected model with $N=10$ qubits. Classical BM will have $N + N (N-1)/2 =  N (N+1)/2$ trainable parameters $(b_a,w_{ab})$. The Hamiltonian of QBM has the form (\ref{eq:Hamiltonian}) where we restrict all $\Gamma_a$ to be the same ($=\Gamma$).  In order to understand the efficiency of QBM in representing data we will train the exact log-likelihood  using (\ref{pL}) and treat $b_a,w_{ab}$, and $\Gamma$ as trainable parameters. This can be done  using exact diagonalization for small systems. We will also perform training of the bound $\tilde{\cal L} $ using (\ref{eq:bound_gradient}) treating $(b_a,w_{ab})$ as trainable parameters but fixing $\Gamma$ to some ad-hoc non-zero value $\Gamma=2$. Comparing the training results of QBM with bQBM will give us some idea of the efficiency of training the bound $\tilde{\cal L}$.

\begin{figure}[t]
 \includegraphics[trim = 20mm 0mm 20mm 30mm, width=6.5cm]{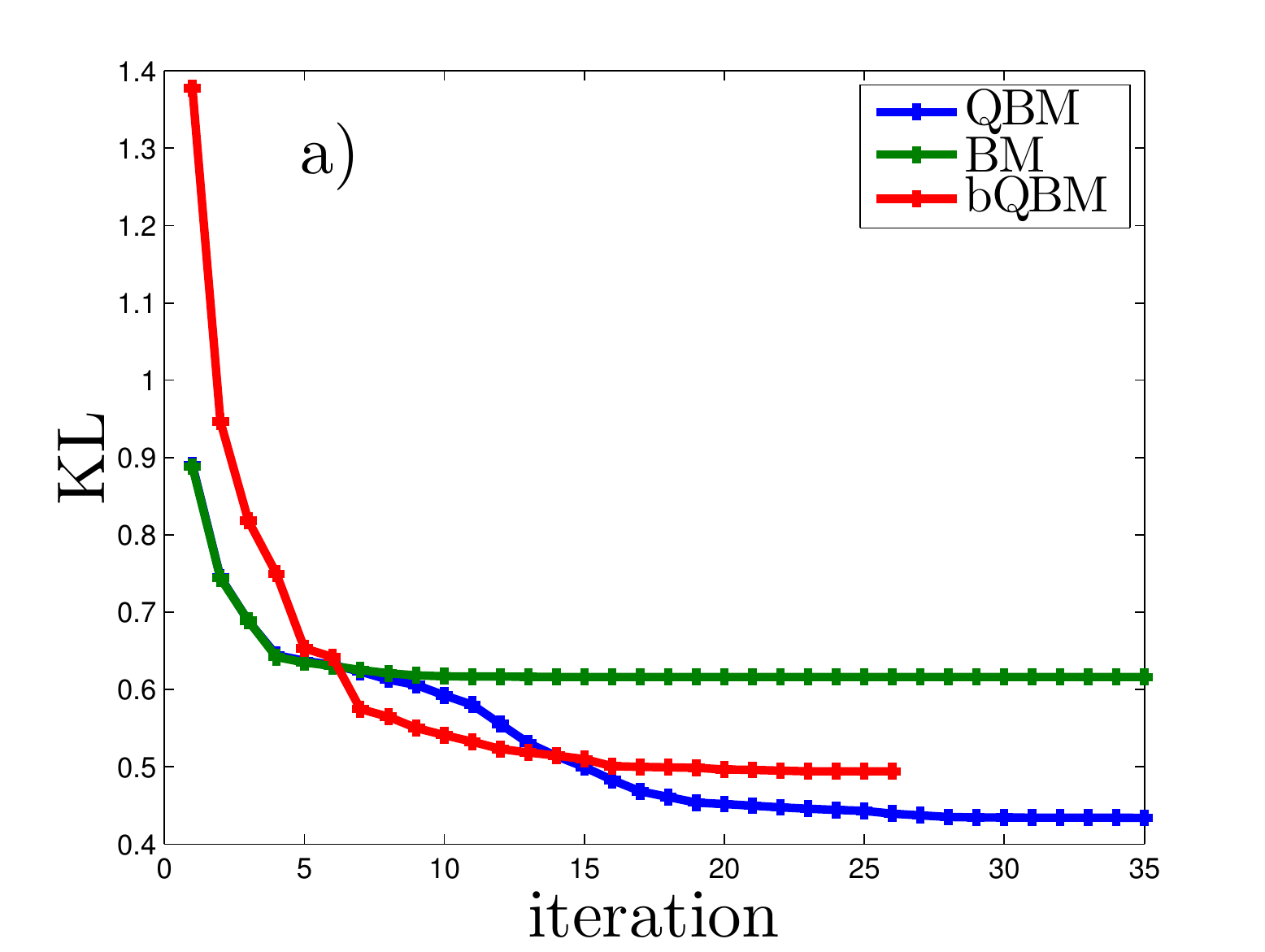}
   \includegraphics[trim = 20mm 0mm 20mm 0mm, width=6.5cm]{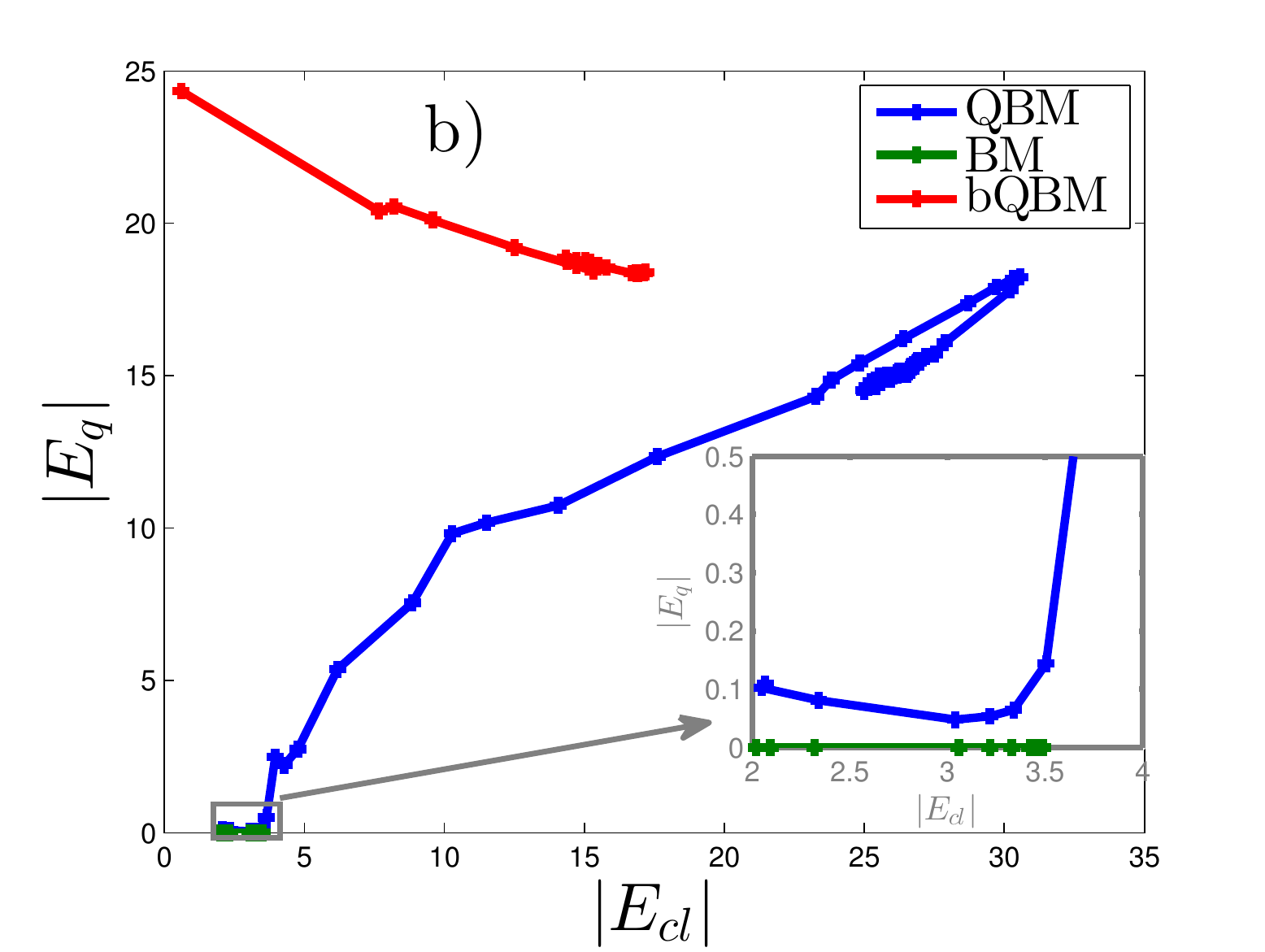}
\caption{Training of a fully visible fully connected model  with $N=10$ qubits  on artificial data from Bernoulli mixture  model (\ref{eq:mixture_model}). Training is done using second-order optimization routine BFGS. (a) KL-divergence (\ref{eq:KL_def}) of BM, QBM, bQBM models during training process. Both QBM and bQBM learn to KL values that are lower than that for BM. (b) Classical and quantum average energies (\ref{eq:cl_q_energies}) during training process.}
    \label{fig:fully_visible}
\end{figure}

Since all expectations entering the gradients of log-likelihood are computed exactly, we will use second-order optimization routine BFGS \cite{bfgs}. The results of training BM, QBM and bQBM are given in Fig \ref{fig:fully_visible}a. The x-axis in the figure corresponds to iterations of BFGS that does line search along the gradient. QBM is able to learn the data noticeably better than BM, and bQBM approaches the value close the one for QBM.

In order to visualize the training process, we keep track of the average values of classical and quantum parts of the Hamiltonian during the training

\ba\label{eq:cl_q_energies}
&& E_{cl} = - \left\langle  \sum_{a} b_a \sigma^z_a + \sum_{a,b} w_{ab} \sigma^z_a \sigma^z_b  \right\rangle, \nn
&& E_q =  - \left\langle  \sum_{a} \Gamma_a \sigma^x_a  \right\rangle.
\ea
Fig \ref{fig:fully_visible}b shows the learning  trajectories in the space $(| E_{cl} |, | E_q|)$. BM learns a model with average energy $\approx 3.5$, and KL $\approx 0.62$. One can see that QBM, which starts off with $\Gamma=0.1$, initially lowers $\Gamma$ and learns $(b_a,w_{ab})$ that are close to the best classical result (see the inset). Soon after, QBM increases $\Gamma$ and  $(b_a,w_{ab})$ until it converges to a point with $\Gamma=2.5$ and KL $\approx 0.42$, which is better than classical BM value. Having a fixed transverse field, $\Gamma=2$, bQBM starts with a large $E_q$ and approaches the parameter learned by QBM (although doesn't reach the best value at $\Gamma=2.5$ learned by QBM).

\subsection{Restricted QBM}\label{subsec:RBM}

\begin{figure}[t]
 \includegraphics[trim = 20mm 0mm 20mm 30mm, width=6.5cm]{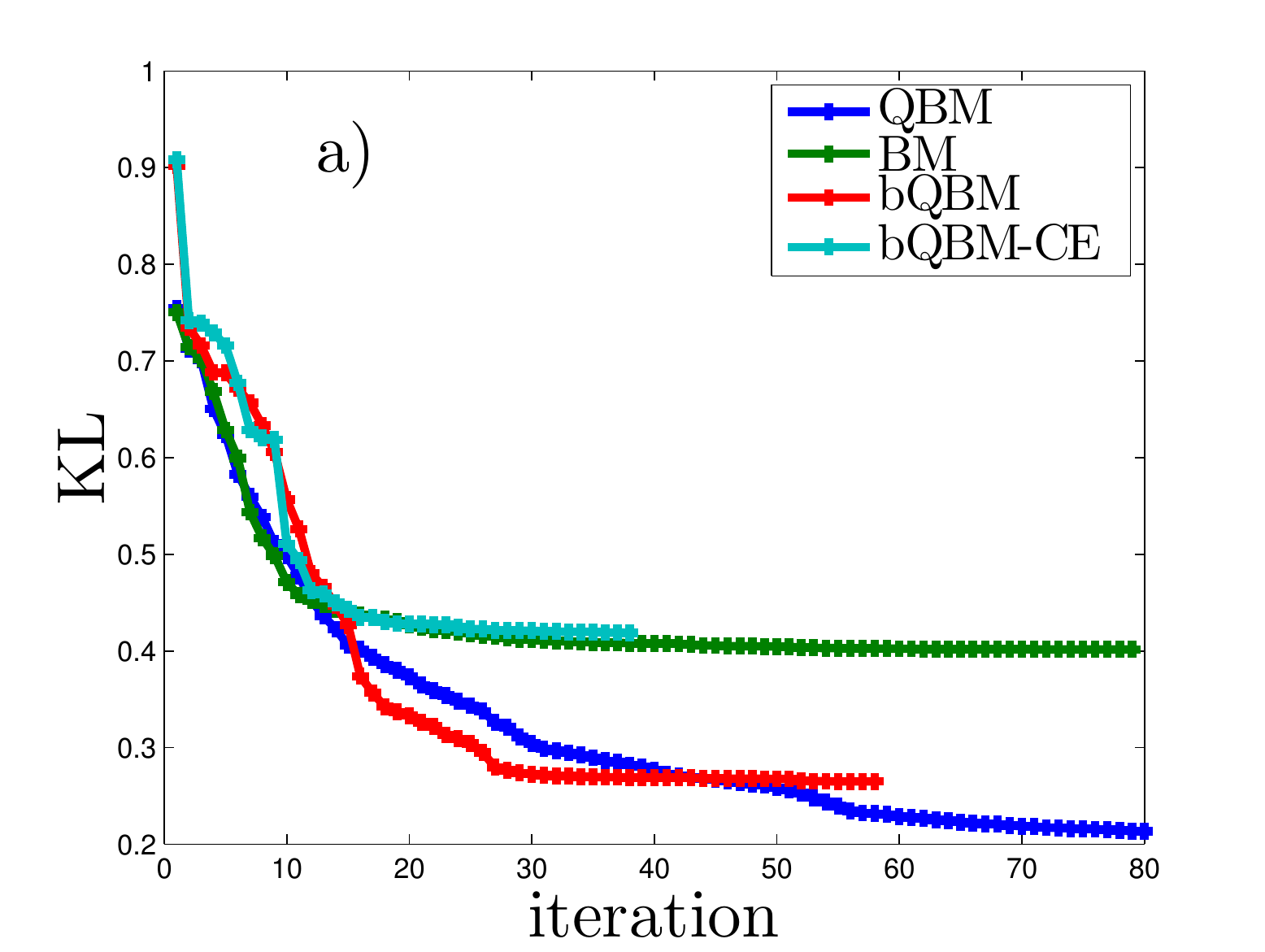}
   \includegraphics[trim = 20mm 0mm 20mm 0mm, width=6.5cm]{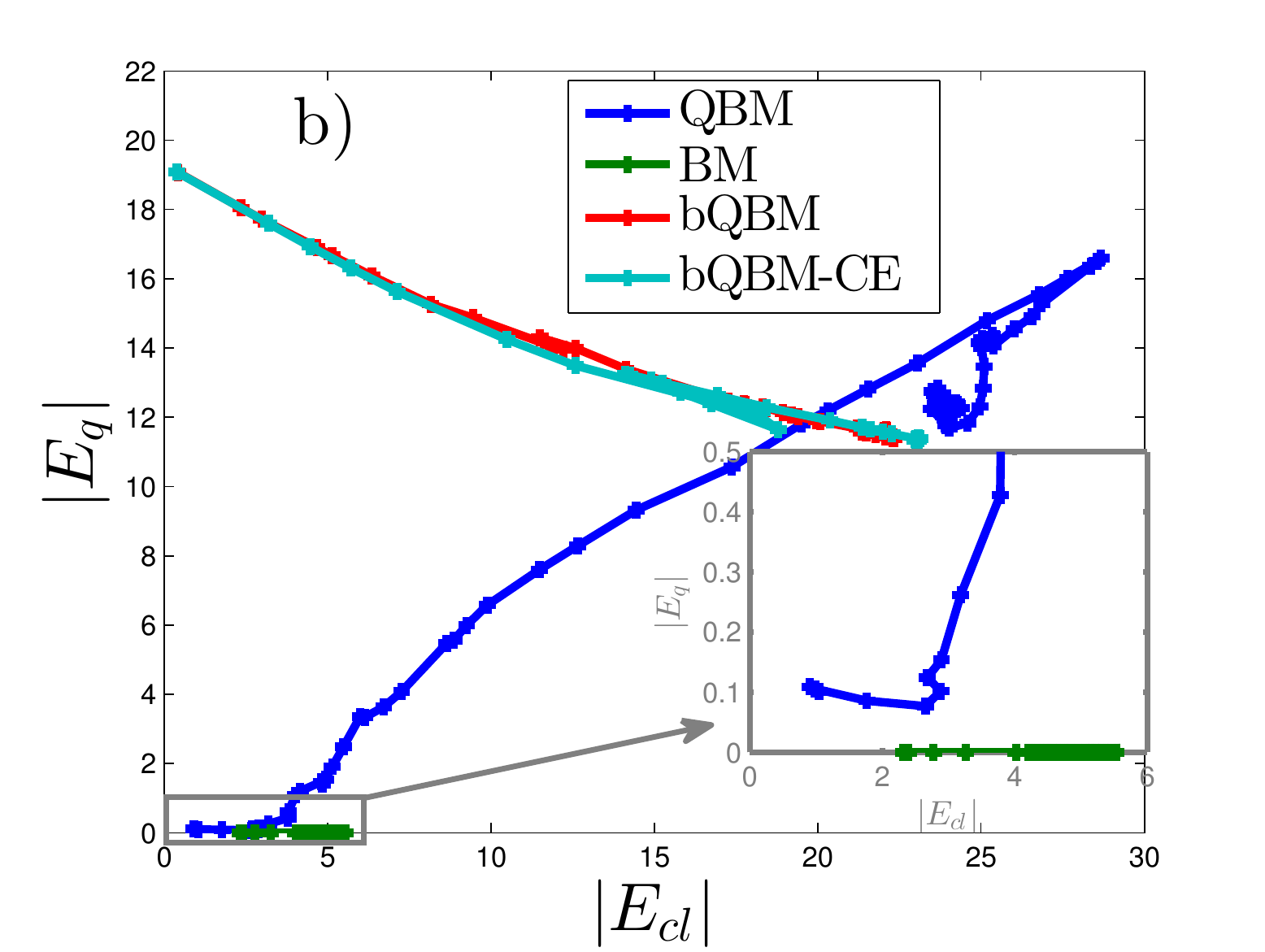}
\caption{Training of a restricted RBM with 8 visible and 2 hidden units on artificial data from Bernoulli mixture  model (\ref{eq:mixture_model}) using second-order optimization routine. (a) KL-divergence (\ref{eq:KL_def}) of different models during training process. Again QBM and bQBM outperform BM, but when positive phase was calculated classically in bQBM, the performance (bQBM-CE curve in the figure) deteriorated and became worse than that for BM (see Section \ref{subsec:RBM} for details). (b) Classical and quantum average energies (\ref{eq:cl_q_energies}) during training process.}
    \label{fig:semi_restricted_rbm}
\end{figure}

We now consider a (semi-) restricted BM discussed in Section \ref{subsec:RQBM}. Our toy model has 8 visible units and 2 hidden units. We allow full connectivity within the visible layer and all-to-all connectivity between the layers. The data is again generated using Eq.~(\ref{eq:mixture_model}) for the visible variables, with $p=0.9$ and $M=8$. We present the results of training in Fig \ref{fig:semi_restricted_rbm}. Similarly to the fully visible model, QBM outperforms BM, and bQBM represents a good proxy for learning quantum distribution.

In order to illustrate the significance of consistent usage of quantum distribution in evaluating the gradients (\ref{eq:qb_gradient}) and (\ref{eq:qw_gradient}), we train bQBM using classical expression instead of (\ref{eq:hidden_exp}) for expectations of hidden units in the positive phase. The resulting machine (bQBM-CE in Fig.~\ref{fig:semi_restricted_rbm}) learns worse than purely classical BM because the two terms in gradient expressions are evaluated inconsistently.

\subsection{Generative supervised learning}

\begin{figure}[t]
 \includegraphics[trim = 20mm 0mm 20mm 30mm, width=6.5cm]{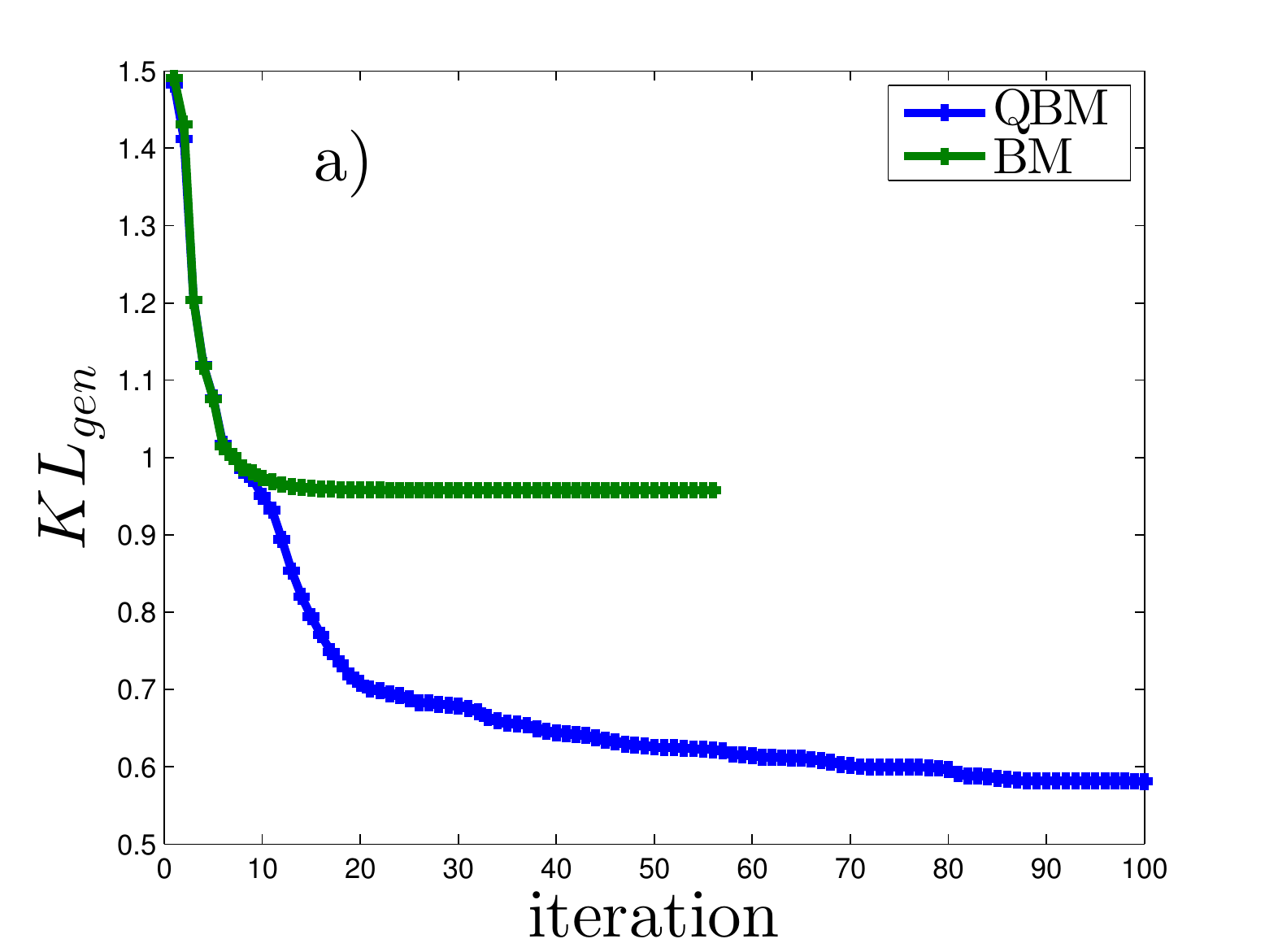}
 \includegraphics[trim = 20mm 0mm 20mm 0mm, width=6.5cm]{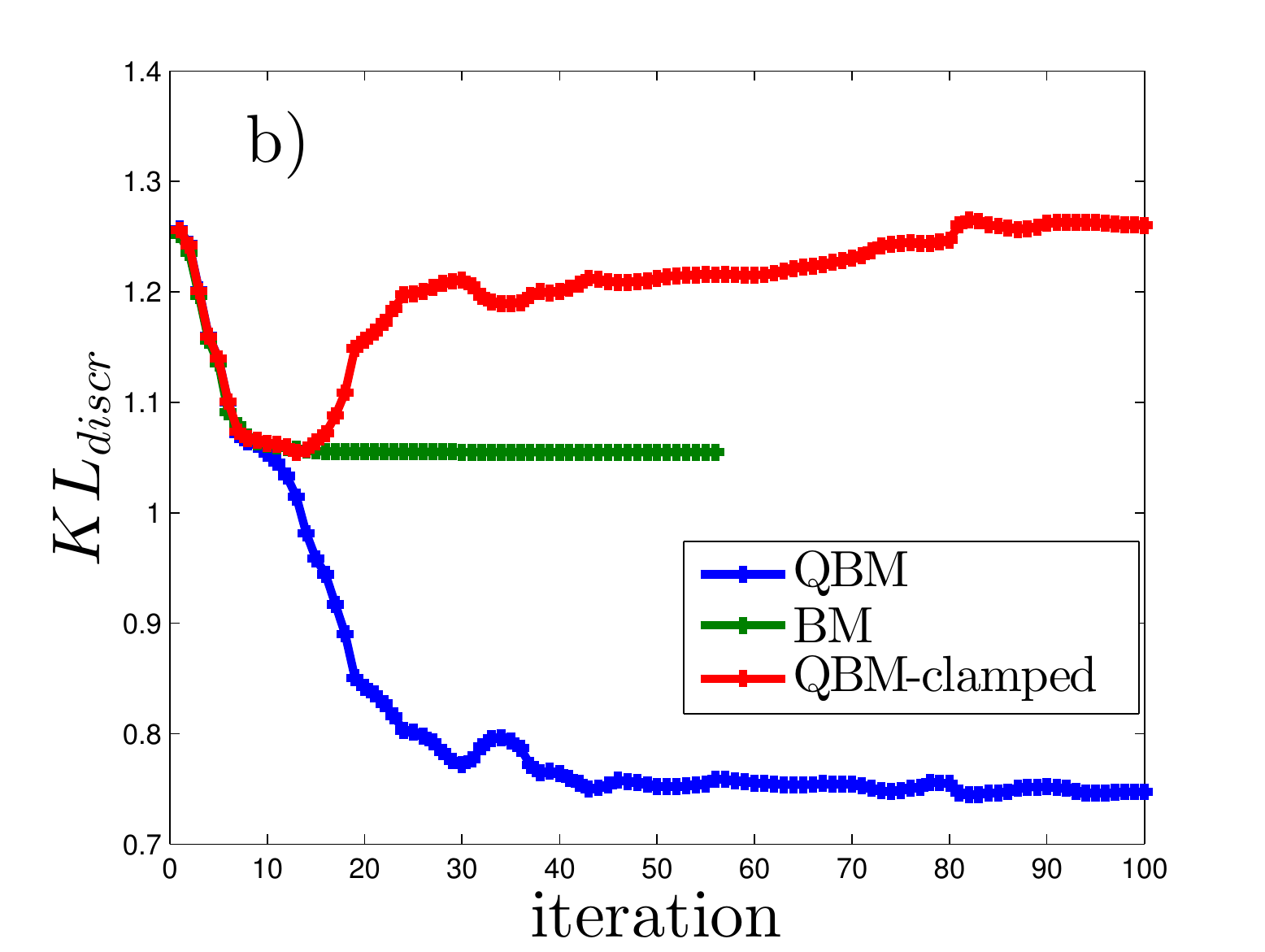}
\caption{Supervised learning using fully visible fully connected  model with $N=11$ qubits divided into 8 inputs and 3 outputs. As our training data we use artificial data from Bernoulli mixture model (\ref{eq:mixture_model}) for inputs and 3-bit binary labels (0 to 7) for outputs. Training is done using second-order optimization routine BFGS. (a) KL-divergence of joint distribution (\ref{eq:L_gen}) of BM, QBM models during training process. Once again QBM learns the distribution better than BM. (b) KL-divergence of conditional distribution during the same training, for BM, QBM models using (\ref{eq:L_discr}), and for clamped QBM (QBM-clamped) using (\ref{eq:conditional_prob_clamped}). The conditional  distribution is also learned better by QBM than BM, but the clamped QBM distribution is very different from the conditional one and give a KL-divergence much higher than the classical BM.}
    \label{fig:generative_learning}
\end{figure}

We consider a supervised learning example with 8 inputs and 3 outputs with full connectivity between all units. For the training set we again used the multi-modal distribution  (\ref{eq:mixture_model}) over ${\bf x}$, with $M=8$ and $p=0.9$, and set the label ${\bf y}$ for each mode to be a 3-bit binary number from 0 to 7 \footnote{This choice was made to keep the number of qubits small to allow for exact diagonalization.}. Both BM and QBM are trained to learn the loss function (\ref{eq:L_gen}). Our goal is to check whether $P^{\rm clamped}_{\bf y|x}  \approx P_{\bf y|x}$, when training QBM in this generative setup. In Fig \ref{fig:generative_learning}a, we plot KL-divergence based on the generative log-likelihood (\ref{eq:L_gen}) for both classical BM and QBM. It is clear that QBM has trained with better KL-divergence than classical BM. In \ref{fig:generative_learning}b we plot the KL-divergence based on the discriminative log-likelihoods  (\ref{eq:L_discr}), evaluated with conditional probabilities $P_{\bf y|x}$ and clamped probabilities $P^{\rm clamped}_{\bf y|x}$. One can see that although QBM is trained with the joint probability distribution, the conditional distribution is also learned better than BM. The clamped probability distribution, on the other hand, starts very close to the conditional distribution at the beginning of the iterations, when QBM and BM are closed to each other. But as the transverse field in QBM starts to grow, the clamped distribution deviates from the conditional one and its KL-divergence grows to a value much worse than the classical BM value. This shows that even for such a small example the clamped distribution can be very different from the true conditional distribution.

\section{QBM with a quantum annealing processor}

Recent developments in manufacturing quantum annealing processors have made it possible to experimentally test some of the quantum machine learning ideas. Up to now many experiments have confirmed the existence of quantum phenomena in such processors \cite{Harris10,Johnson11,Boixo13,BoixoNP14,Boixo15}, which includes entanglement \cite{Lanting14}. A quantum annealing processor implements the time-dependent Hamiltonian
 \be
 {\cal H}(s) = - A(s)\sum_{a}\sigma^x_a + B(s)[\sum_{a}h_i\sigma^z_a {+} \sum_{a,b}
 J_{ab}\sigma^z_a\sigma^z_b ], \label{HP}
 \ee
where $s = t/t_a$, $t$ is time, $t_a$ is the annealing time, $h_i$ and $J_{ij}$ are tuneable dimensionless parameters, and $A(s)$ and $B(s)$ are monotonic functions, with units of energy, such that $A(0) \gg B(0) \approx 0$ and $B(1) \gg A(1) \approx 0$. As discussed in Ref.~\cite{Amin15}, an open system quantum annealier follows quasistatic evolution following equilibrium distribution, $\rho = Z^{-1}e^{-\beta {\cal H}(s)}$, up to a point where the dynamics become too slow to establish equilibrium. Here, $\beta = (k_BT)^{-1}$ with $T$ being the temperature and $k_B$ being the Boltzmann constant. The system will then deviate from the equilibrium distribution and soon after the dynamics will freeze (see Fig.~2c in Ref.~\cite{Amin15} and the related discussion).

In general, a quantum annealer with linear annealing schedule $s=t/t_a$ does not return a Boltzmann distribution. However, as argued in Ref.~\cite{Amin15}, if the dynamical slow-down and freeze-out happen within a short period of time during the annealing, then the final distribution will be close to the quantum Boltzmann distribution of (\ref{HP}) at a single point $s^*$, called the freeze-out time. In such a case, the quantum annealer with linear annealing schedule will provide approximate samples from the Boltzmann distribution corresponding to the Hamiltonian ${\cal H}(s^*)$. Moreover, if $A(s^*)$ happens to be small enough such that the quantum eigenstates at $s^*$ are close to the classical eigenstates, then the resulting Boltzmann distribution will be close to the classical Boltzmann distribution. In such a case, the quantum annealer can be used as an approximate classical Boltzmann sampler for training a BM, as was done in \cite{Adachi15,Alejandro15}. Unfortunately, not all problems have a narrow freeze-out region and $A(s^*)$ is not always small. If the freeze-out does not happen in a narrow region, then the final probability distribution will depend on the history within this region and will not correspond to a Boltzmann distribution at any particular point. This would limit the applicability of using quantum annealer for Boltzmann sampling.

In principle, it is possible to controllably freeze the evolution at a desired point, $s^*$, in the middle of the annealing and readout the qubits. One way to do this is via a nonuniform $s(t)$ which anneals slowly at the beginning up to $s^*$ and then moves very fast (faster than all dynamics) to the end of annealing. An experimental demonstration of such controlled sampling was done in \cite{Dickson13} for a specially designed 16 qubit problem. If $s^*$ lies in the region where the evolution is still quasistatic, the quantum annealer will provide samples from the Boltzmann distribution of Hamiltonian (\ref{HTI}), with
 \ba
 \Gamma_a &=& \Gamma = \beta A(s^*), \\ 
 b_a &=& \beta B(s^*) h_a, \\ 
 w_{ab} &=& \beta B(s^*) J_{ab}.
 \ea
Since $h_a$ and $J_{ab}$ are tunable parameters, if one can control the freeze-out point $s^*$, then all the dimensionless parameters in (\ref{HTI}), i.e., $\Gamma, b_a, w_{ab}$, can be tuned and therefore the quantum annealer can be used for training a QBM.

The applicability of the controlled sampling technique used in \cite{Dickson13} is limited by how fast the second part of the annealing can be done, which is ultimately determined by the bandwidth of the filters that bring electrical signals to the chip. Because of this, such a technique is only applicable to specially designed problems that have very slow dynamics. With some modifications to the current hardware design, however, such techniques can become possible for general problems relevant to QBM in the near future.

\section{Conclusion}

We have examined the possibility of training a quantum Boltzmann machine (QBM), in which the classical Ising Hamiltonian is augmented with a transverse field. Motivated by the success of stochastic gradient descent in training classical Boltzmann machines, one may wish to use a similar technique to optimize the log-likelihood of the QBM. However, unlike the classical BM, for which the gradients of the log-likelihood can be estimated using sampling, the existence of a transverse field in the QBM makes the gradient estimation nontrivial. We have introduced a lower bound on the log-likelihood, for which the gradient can be estimated using sampling. We have shown examples of QBM training through maximizing both the log-likelihood and its lower bound, using exact diagonalization, and compared the results with classical BM training. We have shown small-size examples in which QBM learned the data distribution better than BM. Whether QBM can learn and generalize better than BM at larger sizes are questions that need to be answered in future works.

Our method is different from other existing quantum machine learning proposals  \cite{Lloyd13,Lloyd14,Weibe14,Neven08,Neven08b,Neven09,Pudenz11,deFreitas11,Denchev12,Dumoulin13,Babbush14,Adachi15,Alejandro15}, because quantum mechanics is not only used to facilitate the training process, as in other proposals, but also is exploited in the model. In other words, the probabilistic model, i.e., the quantum Boltzmann distribution, that we use in our QBM is different from any other models that have been studied in the machine learning community. Therefore, the potential of the model for machine learning is unexplored.

We should mention that the similarity between BM and QBM training may not hold in all situations. For example, as we have shown in Sec. \ref{GenLearning}, sampling from a conditional distribution cannot be performed by clamping in QBM, as it is commonly done in classical BM. The two models may also differ in other aspects. Therefore, careful examination is needed before replacing BM with QBM in existing machine learning techniques.

Finally, we have discussed the possibility of using a quantum annealer for QBM training. Although the current commercial quantum annealers like D-Wave are not designed to provide quantum Boltzmann samples, with minor modifications to the hardware design, such a feature can become available. This would open new possibilities in both quantum information processing and machine learning research areas.

\section*{Acknowledgement}

We are grateful to Ali Ghodsi, Firas Hamze, William Macready, Anatoly Smirnov, and Giacomo Torlai for fruitful discussions. This research was partially supported by a Natural Sciences and Engineering Research Council of Canada (NSERC) Engage grant.


\begin{thebibliography}{1}

\bibitem{MLScience} M.I. Jordan and T.M. Mitchell, Machine learning: Trends, perspectives, and prospects, Science {\bf 349}, 255 (2015).

\bibitem{Bishop} C.M. Bishop, Pattern Recognition and Machine Learning, Springer 2006.

\bibitem{Lloyd13} S. Lloyd, M. Mohseni, P. Rebentrost, Quantum algorithms for supervised and unsupervised machine learning, eprint:  arXiv:1307.0411.

\bibitem{Lloyd14} P. Rebentrost, M. Mohseni, S. Lloyd, Quantum support vector machine for big data classification, Phys. Rev. Lett. {\bf 113}, 130503 (2014).

\bibitem{Weibe14} N. Wiebe, A. Kapoor, and K.M. Svore, Quantum Deep Learning, eprint: arXiv:1412.3489.

\bibitem{Neven08} H. Neven, G. Rose, W.G. Macready, Image recognition with an adiabatic quantum computer I. Mapping to quadratic unconstrained binary optimization, arXiv:0804.4457.

\bibitem{Neven08b} H. Neven, V.S. Denchev, G. Rose, W.G. Macready, Training a Binary Classifier with the Quantum Adiabatic Algorithm, arXiv:0811.0416.

\bibitem{Neven09} H. Neven, V.S. Denchev, G. Rose, W.G. Macready, Training a Large Scale Classifier with the Quantum Adiabatic Algorithm, arXiv:0912.0779.

\bibitem{Pudenz11} K.L. Pudenz, D.A. Lidar, Quantum adiabatic machine learning, arXiv:1109.0325.

\bibitem{deFreitas11} M. Denil and N. de Freitas, Toward the implementation of a quantum RBM, NIPS*2011
Workshop on Deep Learning and Unsupervised Feature Learning.

\bibitem{Denchev12} V.S. Denchev, N. Ding, S.V.N. Vishwanathan, H. Neven, Robust Classification with Adiabatic Quantum Optimization, arXiv:1205.1148.

\bibitem{Dumoulin13} V. Dumoulin, I.J. Goodfellow, A. Courville, Y. and Bengio, On the Challenges of Physical Implementations of RBMs, AAAI 2014: 1199-1205.

\bibitem{Babbush14}R. Babbush, V. Denchev, N. Ding, S. Isakov, H. Neven, Construction of non-convex polynomial loss functions for training a binary classifier with quantum annealing, arXiv:1406.4203.

\bibitem{Johnson11} M.W. Johnson, M.H.S. Amin, S. Gildert, T. Lanting, F. Hamze, N. Dickson, R. Harris, A.J. Berkley, J. Johansson, P. Bunyk, E.M. Chapple, C. Enderud, J.P. Hilton, K. Karimi, E. Ladizinsky, N. Ladizinsky, T. Oh, I. Perminov, C. Rich, M.C. Thom, E. Tolkacheva, C.J.S. Truncik, S. Uchaikin, J. Wang, B. Wilson, and G. Rose, Quantum Annealing with Manufactured Spins, Nature {\bf 473}, 194 (2011).

\bibitem{Adachi15} S.H. Adachi, M.P. Henderson, Application of Quantum Annealing to Training of Deep Neural Networks, eprint: arXiv:1510.06356.

\bibitem{Alejandro15} M. Benedetti, J. Realpe-G�mez, R. Biswas, A. Perdomo-Ortiz, Estimation of effective temperatures in a quantum annealer and its impact in sampling applications: A case study towards deep learning applications, arXiv:1510.07611.

\bibitem{Hinton1983} G. E. Hinton, T. J. Sejnowski, Optimal perceptual inference, CVPR 1983.

\bibitem{Hinton2006} G. E. Hinton, S. Osindero, Y-W. Teh, A fast learning algorithm for deep belief nets, Neural Comput. {\bf 18}, 1527--1554 (2006).

\bibitem{Sejnowski1986} T. J. Sejnowski, Higher-order Boltzmann machines, AIP Conference Proceedings 151: Neural Networks for Computing (1986).

\bibitem{Salakhutdinov2009} R. Salakhutdinov, G. E. Hinton, Deep Boltzmann machines, AISTATS 2009.

\bibitem{Ranzato2010} M. Ranzato, G. E. Hinton, Modeling pixel means and covariances using factorized third-order Boltzmann machines, CVPR 2010.

\bibitem{Memisevic2010} R. Memisevic, G. E. Hinton, Learning to represent spatial transformation with factored higher-order Boltzmann machines, Neural Comput. {\bf 22}, 1473--1492 (2010).

\bibitem{Golden65} S. Golden, Lower bounds for the Helmholtz function,
    Phys. Rev., {\bf 137}, B1127 (1965).

\bibitem{Thompson65}  C.J. Thompson, Inequality with applications in statistical mechanics,
    J. Math. Phys. {\bf 6}, 1812 (1965).

\bibitem{bfgs} https://en.wikipedia.org/wiki/Broyden-Fletcher-Goldfarb-Shanno\_algorithm.

\bibitem{semirestricted} S. Osindero and G.E. Hinton, Modeling image patches with a directed hierarchy of Markov random fields, Advances in neural information processing systems (2008).

\bibitem{Harris10} R. Harris {\em et al.}, Experimental Investigation of an Eight Qubit Unit Cell in a Superconducting Optimization Processor, Phys. Rev. B {\bf 82}, 024511 (2010).

\bibitem{Boixo13} S. Boixo, T. Albash, F. M. Spedalieri, N. Chancellor, and D.A. Lidar,
Experimental Signature of Programmable Quantum Annealing, Nat. Commun. {\bf 4}, 2067 (2013).

\bibitem{BoixoNP14} S. Boixo, T.F. R{\o}nnow, S.V. Isakov, Z. Wang, D. Wecker, D.A. Lidar, J.M. Martinis, M. Troyer, Nature Phys. {\bf 10}, 218 (2014).

\bibitem{Boixo15} S. Boixo, V. N. Smelyanskiy, A. Shabani, S. V. Isakov, M. Dykman, V. S. Denchev, M. Amin, A. Smirnov, M. Mohseni, and H. Neven, eprint arXiv:1502.05754, long version: arXiv:1411.4036 (2014).

\bibitem{Lanting14} T. Lanting {\em et al.}, Phys. Rev. X, {\bf 4}, 021041 (2014).

\bibitem{Amin15} M.H. Amin, Searching for quantum speedup in quasistatic quantum annealers, Phys. Rev. A {\bf 92} 052323, (2015).

\bibitem{Dickson13} N.G. Dickson, M.W. Johnson, M.H. Amin, R. Harris, F. Altomare, A. J. Berkley, P. Bunyk, J. Cai, E. M. Chapple, P. Chavez, F. Cioata, T. Cirip, P. deBuen, M. Drew-Brook, C. Enderud, S. Gildert, F. Hamze, J.P. Hilton, E. Hoskinson, K. Karimi, E. Ladizinsky, N. Ladizinsky, T. Lanting, T. Mahon, R. Neufeld, T. Oh, I. Perminov, C. Petroff, A. Przybysz, C. Rich, P. Spear, A. Tcaciuc, M.C. Thom, E. Tolkacheva, S. Uchaikin, J. Wang, A. B. Wilson, Z. Merali, and G. Rose, Thermally assisted quantum annealing of a 16-qubit problem, Nature Commun. {\bf 4} 1903, (2013).


\end{thebibliography}
\end{document}